\documentclass[aps,pra,twocolumn,floatfix,superscriptaddress]{revtex4-1}
               
\usepackage[latin1]{inputenc}
\usepackage[pdftex]{graphicx}
\usepackage{soul,color}
\usepackage{textcomp}
\usepackage{hyperref}
\usepackage{amssymb}
\usepackage{amsmath}
\usepackage{dsfont}
\usepackage{mathrsfs}
\usepackage{ifthen}   
\usepackage{xifthen} 

\usepackage[normalem]{ulem}

\newcommand{\fig}[2][]{%
\ifthenelse{\isempty{#1}}
{Fig.~\ref{#2}}
{Fig.~\ref{#2}(#1)}
}

\makeatletter
\newsavebox{\@brx}
\newcommand{\llangle}[1][]{\savebox{\@brx}{\(\m@th{#1\langle}\)}%
	\mathopen{\copy\@brx\mkern2mu\kern-0.9\wd\@brx\usebox{\@brx}}}
\newcommand{\rrangle}[1][]{\savebox{\@brx}{\(\m@th{#1\rangle}\)}%
	\mathclose{\copy\@brx\mkern2mu\kern-0.9\wd\@brx\usebox{\@brx}}}
\makeatother

\begin{document}

\title{Tuning the nonlinear dispersive coupling of nanomechanical string resonators}


\author{Katrin Gajo}
\author{Gianluca Rastelli}
\author{Eva M. Weig}
\email[]{eva.weig@uni-konstanz.de}

\affiliation{University of Konstanz, Department of Physics, 78457 Konstanz, Germany}

\date{\today}

\begin{abstract}
We investigate nonlinear dispersive mode coupling between the flexural in- and out-of-plane modes of two doubly clamped, nanomechanical silicon nitride string resonators. As the amplitude of one mode transitions from the linear response regime into the nonlinear regime, we find a frequency shift of two other modes. The resonators are strongly elastically coupled via a shared clamping point and can be tuned in and out of resonance dielectrically, giving rise to multimode avoided crossings. When the modes start hybridizing, their polarization changes. This affects the nonlinear dispersive coupling in a non-trivial way. We propose a theoretical model to describe the dependence of the dispersive coupling on the mode hybridization.
\end{abstract}

\maketitle
%
%
%
%
%
%
%
\section{Introduction}
\label{sec:1}
Nonlinear mechanical mode coupling affects the dynamics of driven systems of coupled nanomechanical resonators, even in situations where the modes are tuned far from resonance or from internal resonances~\cite{Mangussi2016,Vinante2014}. 
It is apparent as a dispersive eigenfrequency shift of the coupled mode which depends on the oscillation amplitude of the driven mode.
Nonlinear mode coupling has been demonstrated between the harmonics of doubly clamped beams or strings, between the orthogonal flexural modes of a singly clamped nano- or microcantilever, and between the vibrational modes of a torsional resonator and one of its suspension springs~\cite{Westra,Avoort2010,Lulla,Matheny,Vinante2014,Cadeddu2016,Dong2018,Mathew2018}.

Here we present nonlinear dispersive mode coupling between the fundamental flexural modes of two adjacent, intrinsically coupled nanomechanical resonators. The resonators under investigation are doubly clamped, strongly pre-stressed silicon nitride string resonators with fundamental eigenfrequencies in the range of $6$\,MHz. A simple theoretical model is developed to describe the behavior of the coupled nonlinear system under eigenfrequency tuning. The model reveals that both nonlinear coefficients, the Duffing nonlinearity and the dispersive coupling, depend on the hybridization of the underlying modes. The prospect of controlling the nonlinear coefficients of a system of two or more resonators may be of interest for future applications of nonlinear nanomechanical networks.

%
%
%
%
%
\section{Actuation and detection}
\label{sec:2}
\begin{figure}[t!]
\centering
\includegraphics[scale=1]{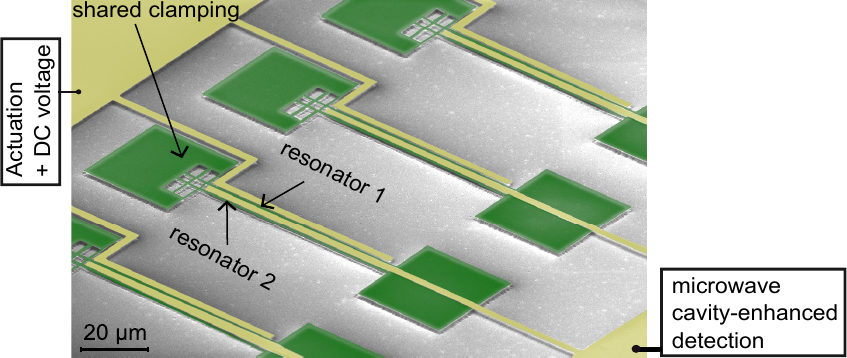}
\caption[]{\label{fig1} (color online) Scanning electron micrograph (false color) of a series of resonator pairs. 
Resonator 1 is dielectrically controlled via a pair of gold electrodes. A RF drive tone and a DC voltage $U_{dc}$ are applied to the outer (upper) electrode. 
A microwave cavity is connected to the inner (lower) electrode situated between the two resonators for cavity-assisted displacement detection of the modes of resonator 1.} 
\end{figure}

The two freely-suspended silicon nitride strings forming the resonator pair are depicted in Fig.~\ref{fig1}.
They share one clamping point, which has been engineered to enable strong linear mechanical coupling between the two string resonators~\cite{Gajo}.
The strings have a nominal length of 55\,$\mu$m, a width of 180\,nm and a thickness of 100\,nm.
The mechanical (flexural) modes have Q-factors in the range of 200\,000 as a result of the high tensile stress (1.46\,GPa) in the silicon nitride nanostrings.
One of the two resonators (resonator 1 in Fig.~\ref{fig1}) is dielectrically controlled~\cite{Unterreithmeier_actuation,Rieger_tuning}.
Actuation is achieved by applying a RF drive tone as well as a DC voltage $U_{dc}$ to one of the electrodes. At the same time, the DC voltage allows to tune the resonator's eigenfrequencies.
The second gold electrode is connected to a $\lambda/4$ microwave cavity for cavity-assisted, 
dielectric displacement detection \cite{Faust_Mikrowelle}. 
To minimize dielectric crosstalk, the RF and DC voltages are applied to the outer electrode, whereas the microwave-cavity-assisted displacement detection is performed on the inner electrode situated between the two string resonators.
All measurements are performed at a pressure below $10^{-4}$ mbar and at room temperature. 
%
%
%
%
%
%
%
%
%
%
%
%
%
%
\section{Characterization in the linear regime}
\label{sec:3}
\begin{figure}[t!]
\centering
\includegraphics[scale=1]{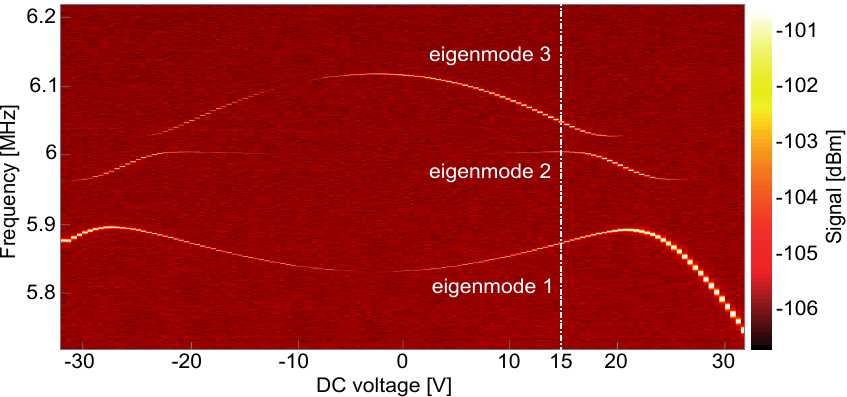}
\caption[]{\label{fig2} 
(color online) Frequency response of hybridized eigenmodes 1 (lowest frequency branch), 2 (intermediate frequency) 
and 3 (highest frequency branch), showing multimode avoided crossings between -30\,V and -20\,V and between 15\,V and 28\,V. 
The white dash-dotted line indicates a DC voltage of 15\,V used for the measurements shown in Fig.~\ref{fig4}. } 
\end{figure}

The two lowest flexural modes of a nanostring correspond to its fundamental out-of-plane (OP) and in-plane mode (IP).
For the case of two nanostrings, one has to deal a priori with four bare, intrinsic fundamental modes 
corresponding to the out-of-plane and in-plane mode of the first resonator (OP1 and IP1) 
and the out-of-plane and in-plane mode of the second resonator (OP2 and IP2). \\
By varying the DC voltage we tune the intrinsic frequencies of the bare modes of resonator 1 such that two (or more) of the four modes become resonant.
In this case, the intrinsic modes hybridize owing to linear inter-string and intra-string mode coupling~\cite{Gajo}.
This linear coupling gives rise to the hybridized eigenmodes of the system.
Figure~\ref{fig2} shows the frequency response of the system as a function of the DC voltage (from -32\,V to 32\,V) applied to the outer electrode.
The three lines correspond to the eigenfrequencies of the hybridized eigenmodes of the system formed by OP1, IP1 and OP2, denoted eigenmode 1, 2 and 3.
The frequency of the bare IP2 mode lies well above all other modes and does not hybridize with the other modes.
Therefore mode IP2 will be disregarded in the following.
Multimode avoided crossings are visible between -30\,V and -20\,V and between 15\,V and 28\,V. 
The white dash-dotted line indicates a voltage of 15\,V, which is used for the measurements shown in Fig.~\ref{fig4}.\\
%
%
%
%
%
%
%
%
%
We consider the simplest theoretical model formed by three linearly coupled harmonic oscillators (corresponding to the bare modes). 
Theoretically the bare modes are described by 
%
%
%
%
%
%
\begin{equation}
\label{eq_equationofmotion_coupledmodes}
\ddot{v}_{i} 
+ \omega^{2}_{i} v_{i} 
+ \sum_{j \neq i} \kappa_{ij} \left( v_{i}-v_{j} \right) 
+ \gamma_{ii} \, v^{3}_{i}
+  \left( \sum_{i \neq j}  \gamma_{ij}   v^{2}_{j} \right) \,  v_{i} =0
 \, ,
\end{equation}
%
%
%
%
($i,j =1,2,3$), where $v_{i}$ is the vibration amplitude of the bare modes with $i=1$ (OP1), $i=2$ (OP2), $i=3$ (IP1), $\omega_{i}$ is the eigenfrequency of the $i$-th bare mode, and 
$\kappa_{ij}$ describes the linear coupling between the bare modes $i$ and $j$.
The parameters $ \gamma_{ii}$  are the Duffing nonlinearities whereas  $\gamma_{ij} $ $(i\neq j)$ are the dispersive coupling coefficients (with the notation $\gamma_{ij}=\gamma_{ji}$). 
We omit nonlinear interactions of the kind $v_i v_j$ 
on the basis that they are related to a breaking of symmetry in the suspended nanowire and generally negligible. 
Furthermore, we ignore the external drive, the damping and the noise.
We will include these terms in the subsequent sections.

In this section we focus on the linear regime by disregarding, for the moment, the nonlinearities, 

%
%
%
\begin{equation}
\label{eq_equationofmotion_coupledmodes_2}
\ddot{v}_{i} 
+ \omega^{2}_{i} v_{i} 
+ \sum_{j \neq i} \kappa_{ij} \left( v_{i}-v_{j} \right) 
=0
 \, .
\end{equation}
%
%
%
%
%
We introduce a voltage dependence for the intrinsic eigenfrequencies $\omega_{i}$ of the bare modes in the system.
As only resonator 1 is surrounded by the gold electrodes, we expect to control only the modes of resonator 1 by our dielectric actuation and tuning technique.
However, we find experimentally that mode OP2 of resonator 2 is also slightly affected by the DC voltage.
Therefore, we assume $\omega_{i}(U_{dc})/(2\pi) =\omega_{i0}/(2\pi) + c_{i}(U_{dc}-U_{i0})^2+   d_{i}(U_{dc}-U_{i0})^3$  
with $\omega_{i0}$ ($i=$1,2,3) being the (theoretical) intrinsic eigenfrequencies of the bare modes in absence of coupling and vanishing voltage.
We have considered a quadratic tuning factor $c_{i}$ along with a cubic correction $d_{i}$.
To be more general, we have also assumed an offset voltage $U_{i0}$ of the vertices of the tuning parabolas.
A voltage offset can arise as a consequence of a buildup of static polarization.
Additionally we assume the linear coupling $\kappa_{ij}$ to be independent of the DC voltage, neglecting dielectric effects.\\
The equations of motion can be re-written as
%
%
%
%
%
%
\begin{equation}\label{eq_bewegung_matrix}
\ddot{\vec{v}}_{i}=-\Theta\vec{v}_{i} \, ,
\end{equation}
%
%
%
%
with the mode matrix $\Theta$ defined as
%
%
%
%
%
%
\begin{gather}
\label{eigenvaluematrix}
\Theta=
\begin{bmatrix}
\omega^{2}_{1}+\kappa_{12}+\kappa_{13} & -\kappa_{12} & -\kappa_{13} \\
-\kappa_{12} & \omega^{2}_{2}+\kappa_{12}+\kappa_{23} & -\kappa_{23} \\
-\kappa_{13} & -\kappa_{23} & \omega^{2}_{3}+\kappa_{13}+\kappa_{23}\\
\end{bmatrix}
\, . 
\end{gather} 
%
%
%
The eigenvalues $\Omega_{i}$ of the matrix $\Theta$ in Eq. (\ref{eigenvaluematrix}) denote the eigenfrequencies of the three hybridized modes.
They depend on the DC voltage, the DC tuning strength and the eigenfrequencies of the bare modes. 
We used a genetic fit~\cite{Gajo} to obtain the parameters entering in the frequencies $\Omega_{i}$, which we extracted from the experimental data in Fig.~\ref{fig2}.
The result of the fitting yields the following parameters: 
$\kappa_{12}/{(2\pi)}^2=4.138 $\,MHz$^{2}$, 
$\kappa_{13}/{(2\pi)}^2=1.9098$\,MHz$^{2}$, 
$\kappa_{23}/{(2\pi)}^2=2.546$\,MHz$^{2}$, 
$\omega_{10}/(2\pi)=5.8606$\,MHz, 
$\omega_{20}/(2\pi)=5.9897$\,MHz, 
$\omega_{30}/(2\pi)=6.1117$\,MHz,
$U_{10}=0.656$\,V,
$U_{20}=-1.1$\,V,
$U_{30}=0.485$\,V,
$c_{1}/(2\pi)= 165.59$\,Hz/V$^{2}$,
$c_{2}/(2\pi)= 1$\,Hz/V$^{2}$,
$c_{3}/(2\pi)=-282.5$\,Hz/V$^{2}$,
$d_{1}/(2\pi)= 1.368$\,Hz/V$^{3}$,
and $d_{3}/(2\pi)=-2.2$\,Hz/V$^{3}$.

%
%
%
%
%
\begin{figure}[t!]
\centering
\includegraphics[scale=1]{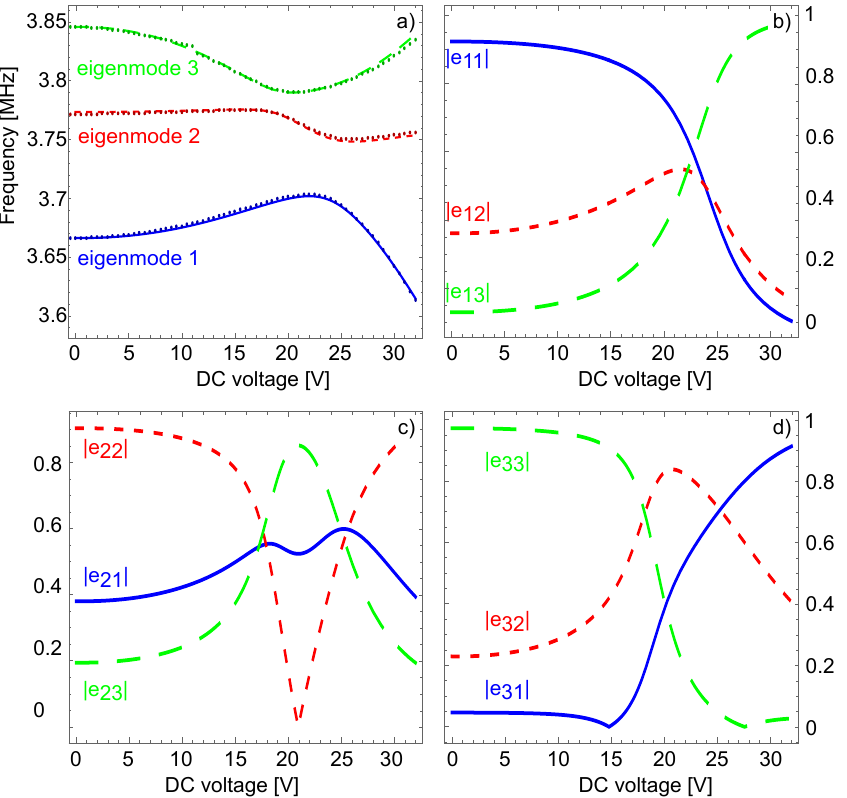}
\caption[]
{\label{fig3} (color online) a)
Frequencies of the hybridized eigenmodes $\Omega_i$ for $i=1$ (blue), $i=2$ (red) and $i=3$ (green). 
Dotted lines represent experimental data, whereas solid or dashed lines represent the results of the fit using the parameters specified in the text.
b), c) and d): components of the normalized eigenvectors $\vec{e}_{1}$, $\vec{e}_{2}$ and $\vec{e}_{3}$ of the three modes.
Each eigenvector has three contributions in the basis formed by the three relevant bare modes of the two 
resonators (OP1, IP1 and OP2).
For example, b) shows the components of eigenmode 1, with $e_{11}$, $e_{12}$, and $e_{13}$ being the contributions of the bare modes OP1, OP2, and IP1, respectively. 
}
\end{figure}
%
%
%
%
In Fig.~\ref{fig3}a we show the three calculated hybridized eigenfrequencies $\Omega_i$ 
(blue solid line, red short dashed line, green long dashed line) based on the above parameters for voltages between 0\,V and 32\,V. As a comparison we show the extracted data from Fig. \ref{fig2} (bue, red and green dotted lines).
As the DC voltage dependence of the eigenfrequencies is (almost) symmetric for positive and negative voltages, 
we further only consider the positive voltage spectrum.\\
Using the transformation 
%
%
%
%
%
%
\begin{equation}
\label{eq:trans}
v_{i}=\sum^{3}_{j=1} \, e_{ji}\left(V_{dc} \right)  \,\,\, q_{j}
\end{equation}
%
%
%
%
we set the variables $q_{j}$ as the vibration amplitude of the hybridized eigenmodes.
Figures ~\ref{fig3}b,c and d show the (voltage dependent)  components $e_{ji}$ of the calculated eigenvectors of the three hybridized eigenmodes, respectively.
One can observe that the three bare modes are always partially hybridized even in the limit of vanishing DC voltage, since $e_{ji}$ remains non-zero even for $j\neq i$.
On the other hand, in the limit $U_{dc} \simeq 0$, each hybridized mode is dominated by the contribution of one particular bare mode, $e_{jj}$, of the system.  
For example,  Fig.~\ref{fig3}b shows that the hybridized eingenmode 1 is dominated by the bare mode OP1
whereas  Fig.~\ref{fig3}c and d show that eigenmodes 2 and 3 are dominated by the contribution of OP2 and IP1, respectively.
This is also deducible from  Fig.~\ref{fig3}a, which shows that eigenmode 1 and 3 both depend strongly 
on the DC voltage which is applied on resonator 1.
On the other hand, mode 2 is almost flat.
This mode has the strongest contribution from the bare mode OP2 (see Fig.~\ref{fig3}c), which is only indirectly affected by the tuning of resonator 1.\\
Finally, the bare modes strongly hybridize in the area of the avoided crossings between 15\,V and 28\,V. 
Note that the microwave cavity-assisted  readout technique is sensitive only to the modes of resonator 1 which is situated between the electrodes. 
Therefore a given hybridized eingenmode can only be detected if it has a sufficiently strong contribution from a bare mode of resonator 1.
%
%
%
%
%
%
%
%
%
\section{Dispersive nonlinear coupling}
\label{sec:4} 
\begin{figure}[t!]
\centering
\includegraphics[scale=1]{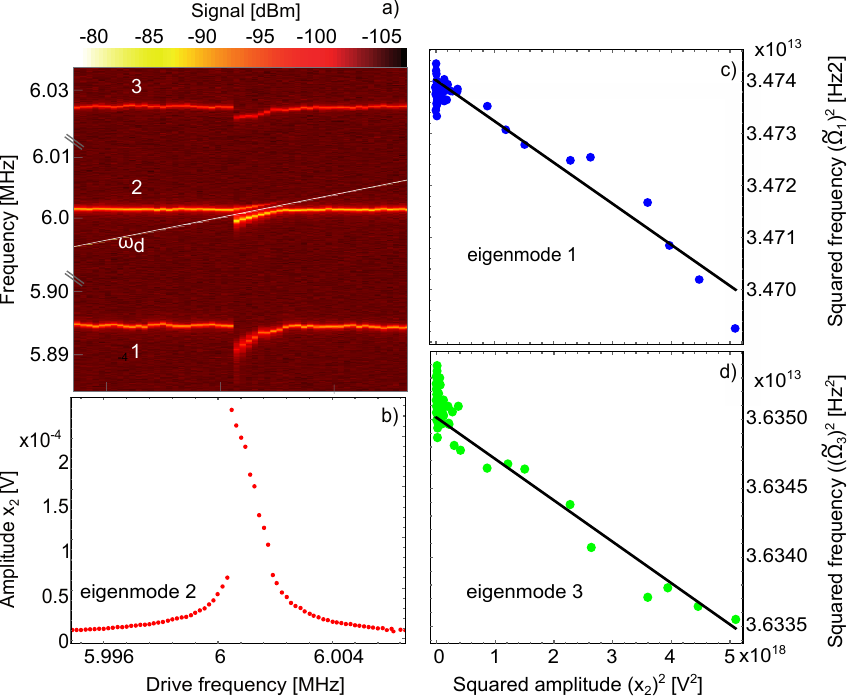}
\caption[]{\label{fig4} (color online) 
a) Power spectra for varying drive frequency at a DC voltage of 15\,V.
The (enhanced) thermal fluctuations of the three hybridized modes 
 $\Omega_{1}$, $\Omega_{2}$ and $\Omega_{3}$ are clearly apparent.
When the drive frequency $\omega_d$ (indicated by white line) approaches the eigenfrequency $\Omega_{2}$, 
this mode is excited and vibrates at the frequency $\omega_d$.
b) Frequency response of the driven eigenmode 2 reflecting the typical Duffing curve. 
c) and d) Dependence of the squared renormalized eigenfrequencies of the two undriven eigenmodes 1 (blue dots) and 3 (green dots)
as a function of the vibration amplitude of the driven eigenmode 2.
The black lines depict the fit of Eq.~(\ref{eq_fit_lambda_a}) and Eq.~(\ref{eq_fit_lambda_b}). 
} 
\end{figure}

We now fix the DC voltage to 15\,V and apply a sinusoidal drive tone $\omega_d$ close to the eigenfrequency $\Omega_{2}$ of the intermediate hybridized eigenmode.
In addition white noise (power of -5dBm,  bandwidth of 7 MHz)  
is applied to enhance the thermal motion of the resonators. 
We sweep the drive frequency $\omega_d$ and measure the power spectrum which is reported 
in Fig.~\ref{fig4}(a).
Far away from resonant drive, i.e. $\omega_d  \neq  \Omega_2$, the three horizontal lines correspond to the (enhanced) thermal motion of the three hybridized eigenmodes 
at the DC voltage 15\,V.
As the drive tone approaches the eigenfrequency,  $\omega_d  \approx  \Omega_2$ (viz. $\Omega_2-\omega_d$ becomes comparable to the linewidth), 
one starts to drive eigenmode 2. 
This is clearly visible in Fig.~\ref{fig4}(b) which displays the measured frequency response of the driven eigenmode 2, namely its vibration amplitude  
as a function of the drive frequency. 
The vibrational state of eigenmode 2 is clearly in the nonlinear regime and the vibration amplitude shows the characteristic shape of the Duffing oscillator 
(with negative, i.e. softening Duffing nonlinearity).
In our high quality resonators operating at room temperature, 
the nonlinear regime of the driven eigenmode manifests in the power spectrum as the formation of two satellite peaks~\cite{Huber2019} around the drive tone (almost a delta peak), clearly resolved in the frequency range  $\omega_d  \approx  \Omega_2$.
\\
Here we focus our attention on the other two undriven hybridized eigenmodes which are apparent 
in the power spectrum as a result of the presence of the noise-enhanced thermal fluctuation.
Figure~\ref{fig4}(a) reveals a clear frequency shift for both undriven eigenmodes 1 and 3, as the drive frequency approaches the eigenfrequency of eigenmode 2.
This occurs when the vibration amplitude of the driven eigenmode strongly increases as we cross the resonant condition $\omega_d  \approx  \Omega_2$.
%
%
%
%
%
%
%
%
%
%
%
%
%
Far away from any internal resonance, the minimal theoretical model that describes the experimental data is given by assuming a dispersive coupling between 
the driven eigenmode, see Eq.~(\ref{eq_equationofmotion_coupledmodes}), and the undriven, fluctuating eigenmodes.
We consider the following equations of motion:
for the driven eigenmode $(\omega_d \approx \Omega_2$) we set 
%
%
%
%
%
%
\begin{equation}
\label{eq:mode_driven}
\ddot{q}_{2}(t) =  -\Omega^{2}_{2} q_{2}(t) -  \Gamma_2 \dot{q}_2(t)  - \tilde{\gamma}_{22} q_2^3(t)  + F_{d} \cos(\omega_{d}t) + \delta F_2(t)  \, , \\
\end{equation}
whereas for the other two eingenmodes  
%
%
%
%
%
%
%
%
\begin{align}
\ddot{q}_{1}(t) &= - \left( \Omega^{2}_{1} +  \tilde{\gamma}_{12} \overline{q^{2}_{2}(t)} \right) q_{1}(t) -  \Gamma_1 \dot{q}_1(t)  + \delta F_1(t) \, ,  \label{eq:mode_notdriven_1}\\
\ddot{q}_{3}(t) & = - \left( \Omega^{2}_{3} +  \tilde{\gamma}_{23} \overline{q^{2}_{2}(t)} \right) q_{3}(t) -  \Gamma_3 \dot{q}_3(t)  + \delta F_3(t) \, . \label{eq:mode_notdriven_2}
\end{align}
%
%
%
%
In  Eq.~(\ref{eq:mode_driven}) and Eqs.~(\ref{eq:mode_notdriven_1},\ref{eq:mode_notdriven_2}) $\Gamma_i$ represents the damping of the $i$-th eigenmode,
and $\delta F_i(t)$ the (enhanced) thermal noise.
The parameter $ \tilde{\gamma}_{22} $ is the Duffing nonlinearity of the driven eigenmode at the DC voltage 15 V.
Since the two undriven eigenmodes are weakly fluctuating only due to the (enhanced) thermal noise, we neglect their effect of nonlinear coupling on the driven eigenmode 
(this term should correspond to a noise effect in the frequency of the driven eigenmode)~\cite{Zhang2015}.
As the driven mode is oscillating at the drive frequency, far away from the eigenfrequency $\Omega_1$ and $\Omega_3$ of the other two eigenmodes,
we have to take only the time average $ \overline{q^{2}_{2}(t)}$ of the driven mode to account for its effect on the other two modes.
We set the latter time average as $\overline{q^{2}_{2}(t)} = (\tilde{\text{c}}_{2}x_2^2) / 2$, 
with $x_2$ the measured signal and $\tilde{\text{c}}_{2}$ a calibration factor. 
Then we can write the renormalized eigenfrequencies of eigenmodes 1 and 3 as a function of the amplitude of the driven eigenmode 2 as
%
%
%
%
%
\begin{align}
\tilde{\Omega}_1^2 &= \Omega_1^2  \, + \, \frac{1}{2}\tilde{\gamma}_{12} \,\tilde{\text{c}}_{2}\,  x_2^2 \, , \label{eq_fit_lambda_a} \\
\tilde{\Omega}_3^2 &= \Omega_3^2  \, + \, \frac{1}{2}\tilde{\gamma}_{23} \,\tilde{\text{c}}_{2}\,  x_2^2  \, . \label{eq_fit_lambda_b}
\end{align}
%
%
%
As mentioned above, the dispersive coupling manifests itself in a frequency shift of one mode, if the vibration amplitude of another mode is changed. \\
Figures \ref{fig4}c and \ref{fig4}d show the shift of the  frequencies $\tilde{\Omega}_1^{2}$ of eigenmode 1 (blue dots) 
and $\tilde{\Omega}_3^{2}$ of eigenmode 3 (green dots) as a function of the amplitude of the driven eigenmode. 
We carry out a fit of Eq.~(\ref{eq_fit_lambda_a}) and Eq.~(\ref{eq_fit_lambda_b}) to the data (black lines) to obtain the dispersive coupling 
(including a calibration factor $\tilde{\text{c}}_{2}$ 
which accounts for the displacement sensitivity of eigenmode 2) 
and find   
$\tilde{\text{c}}_{2} \tilde{\gamma}_{12}=-6.694\cdot 10^{19} \,$Hz$^{2}$/V$^{2}$ 
and 
$\tilde{\text{c}}_{2} \tilde{\gamma}_{23}=-2.368\cdot 10^{19} \, $Hz$^{2}$/V$^{2}$.

%
%
%
%
%
\section{Voltage dependence of Duffing nonlinearity and dispersive coupling}
\label{sec:5}
We repeat the evaluation discussed so far for the other modes, namely by driving either eigenmode 1, 2 or 3 
and detecting the frequency shifts of the respective other eigenmodes.
This allows to extract all dispersive coupling coefficients $\tilde{\text{c}}_{i}\tilde{\gamma}_{ij}$  and $\tilde{\text{c}}_{j}\tilde{\gamma}_{ij}$
as well as Duffing nonlinearities $\tilde{\text{c}}_{i}\tilde{\gamma}_{ii}$ of the eigenmodes (with $i,j=1,2,3$ and $i\neq j$).
Note that we detect different modes, each having a different (and voltage-dependent) calibration factor $\tilde{\text{c}}_{i}$ 
as a result of the mode-polarization dependent displacement sensitivity of our detection scheme.

\begin{figure}[h!]
\centering
\includegraphics[scale=0.33]{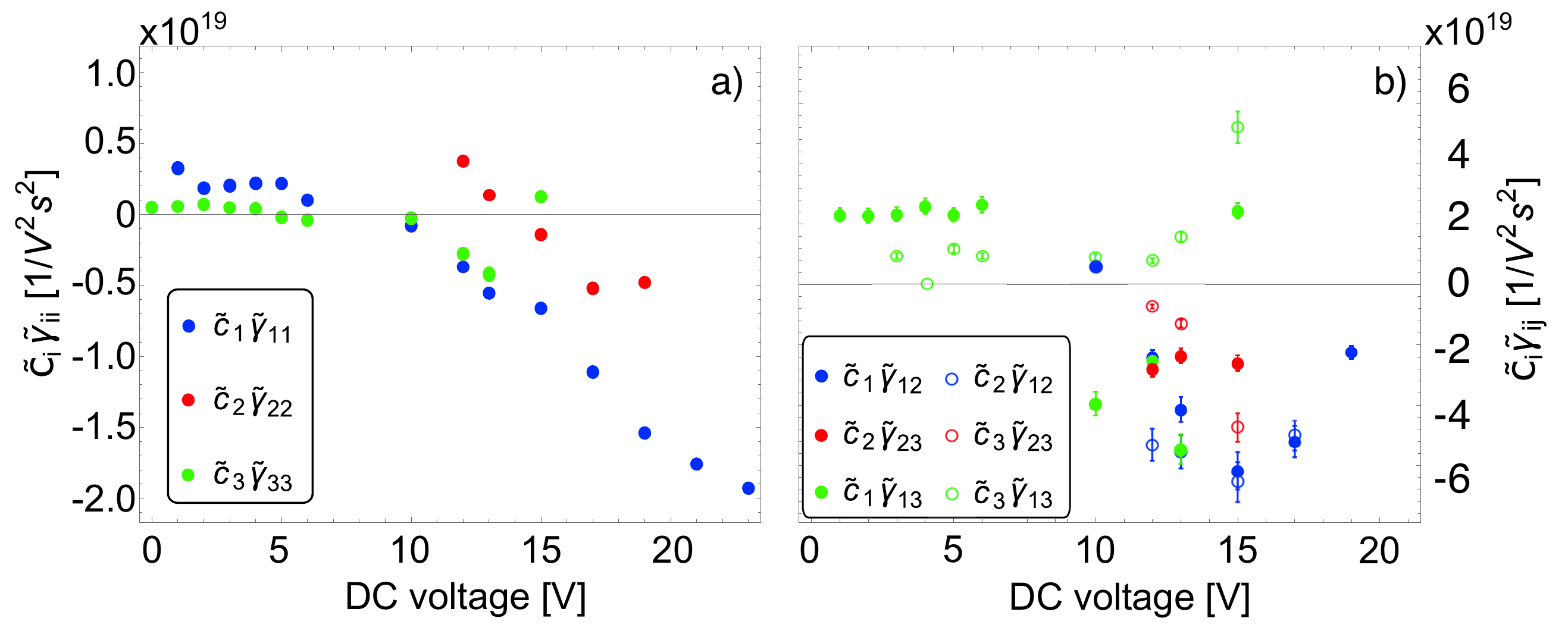}
\caption[]{(color online) 
Nonlinear coefficients (including mode- and voltage-dependent calibration factor) extracted from the experimental data. a) 
Polarization-dependent Duffing nonlinearities 
($\tilde{\text{c}}_{1}\tilde{\gamma}_{11}$: blue dots, $\tilde{\text{c}}_{2}\tilde{\gamma}_{22}$: red dots, $\tilde{\text{c}}_{3}\tilde{\gamma}_{33}$: green dots). 
b) Polarization-dependent dispersive coupling coefficients 
($\tilde{\text{c}}_{1}\tilde{\gamma}_{12}$: blue dots, $\tilde{\text{c}}_{2}\tilde{\gamma}_{12}$: blue circles, 
$\tilde{\text{c}}_{2}\tilde{\gamma}_{23}$: red dots, $\tilde{\text{c}}_{3}\tilde{\gamma}_{23}$: red circles, 
$\tilde{\text{c}}_{3}\tilde{\gamma}_{13}$: green dots, $\tilde{\text{c}}_{1}\tilde{\gamma}_{13}$: green circles)}.
\label{fig5} 
\end{figure}

We repeat the whole procedure for different DC voltages in a range of 0\,V to 22\,V 
to find the DC voltage dependence of the dispersive coupling  
$\tilde{\text{c}}_{i}\tilde{\gamma}_{ij}$, $\tilde{\text{c}}_{j}\tilde{\gamma}_{ij}$
 and $\tilde{\text{c}}_{i}\tilde{\gamma}_{ii}$.
The results are summarized in  Fig.~\ref{fig5}a and Fig.~\ref{fig5}b, which depict the 
experimentally determined Duffing nonlinearities and dispersive coupling coefficients, 
both including the appropriate calibration factor, respectively.
%
%
%
%
%

%
%
%
%
%
\section{Hybridization-dependent Duffing nonlinearity and dispersive coupling}
\label{sec:6}
In this section we present a theoretical model that describes 
the voltage dependence of the Duffing nonlinearities and of the 
dispersive coupling coefficients. 
It is based on the DC-voltage dependent hybridization of the modes, see Fig.~\ref{fig3}.

We start from  
the nonlinear potential of fourth order which couples the three bare modes of the system (OP1, OP2, IP1) 
%
%
%
%
%
%
%
\begin{equation}
\label{eq_potential_0}
\mathcal{V}\left[ \{ v_i \} \right] = 
\frac{1}{4} \sum_{i=1}^{3} \gamma_{ii} v^{4}_{i}
+
\frac{1}{2}
\left(
\gamma_{12}  v^{2}_{1} v^{2}_{2} 
+
\gamma_{13}  v^{2}_{1} v^{2}_{3}
+
\gamma_{23}  v^{2}_{2} v^{2}_{3}  
\right)
 \, . 
\end{equation} 
%
%
%
%
%
with amplitudes $v_i$.
Using the transformation Eq.~(\ref{eq:trans}), 
we change the basis and we express the quartic potential $\mathcal{V}$ of Eq.~(\ref{eq_potential_0}) in 
terms of the hybridized  eingenmodes of the system which are linearly independent. 
The result reads as follows:   
%
%
%
%
\begin{align}
\label{eq_potential_1}
\tilde{\mathcal{V}}  \left[ \{ q_i \} \right]
&= 
\frac{1}{4} \sum_{i=1}^{3} \tilde{\gamma}_{ii} q^{4}_{i}
+
\frac{1}{2}
\left(
\tilde{\gamma}_{12}  q^{2}_{1} q^{2}_{2} 
+
\tilde{\gamma}_{13}  q^{2}_{1} q^{2}_{3}
+
\tilde{\gamma}_{23}  q^{2}_{2} q^{2}_{3}  
\right)
 \nonumber \\
&+  \sum_{i \neq j}   \psi_{ij} q_{i}^{3} q_{j}  + \sum_{ijk}  |\epsilon_{ijk}|  \, \phi_{i} \, q_{i}^{2} q_{j} q_{k}  \, ,
\end{align} 
%
%
%
with the Levi-Civita antisymmetric tensor $ \epsilon_{ijk}$.
The nonlinear potential $\tilde{\mathcal{V}}$ of Eq.~(\ref{eq_potential_1}) for the amplitudes of the hybridized eigenmodes 
contains cubic nonlinearities and even three-body interactions.
However, far from any internal resonances between the frequencies $\Omega_i$ of the hybridized eigenmodes,
the cubic and the three-body interaction terms can be neglected if we apply a single drive frequency. 
In other words, we only consider the Duffing nonlinearity of the driven mode, and the dispersive coupling to 
describe the frequency-shift of the non-driven, thermally activated modes.
%
%
%
%
%
Table \ref{theoretical_summary}  summarizes the notation for the different modes and their interactions.\\
%
%
%
%
%
%
\begin{table}[h!]
\begin{center}
\begin{tabular}{| c | | c | c |}
\hline
					& Bare  modes		&	Hybridized modes					\\
\hline
\hline
Amplitude 			&  $v_i$			&	$q_i $ $ \left( e_{ij}(V_{dc}) \right)$				\\
\hline
Linear interaction		& $\kappa_{ij}$		&		-							\\
\hline
Nonlinear interaction		& $\gamma_{ij}$	& 	$\tilde{\gamma}_{ij}\left(V_{dc} \right)$	\\	
\hline
\end{tabular}
\caption{
Summary of the notations used for the bare intrinsic modes 
and the hybridized eigenmodes.
}
\label{theoretical_summary}
\end{center}
\end{table}
%

In Eq.~(\ref{eq_potential_1}), the Duffing nonlinearities $\tilde{\gamma}_{ii}$
of the  hybridized eigenmodes can be related to the Duffing nonlinearities $\gamma_{ii}$ and the dispersive coupling coefficients $\gamma_{ij}$
of the bare modes via the components of the eigenvectors 
of the hybridized eigenmodes as follows:   
%
%
%
%
%
\begin{equation}
\label{eq_duffing}
\tilde{\gamma}_{ii}
=
\sum_{j}
e^{4}_{ij}
\gamma_{jj}+
\sum_{j \neq k}
e^2_{ij}
e^2_{ik}
\gamma_{jk} \, .
\end{equation}
%
%
%
%
Similarly we obtain the relation for the dispersive coupling coefficients of the hybridized eigenmodes  
%
%
%
%
%
%
\begin{align}
\label{eq_dispersive}
\tilde{\gamma}_{i  j} = & 
6 \sum_{k}
e^{2}_{ik} \, 
e^{2}_{jk} \, 
\gamma_{kk}^{\phantom{2}}
+  
\sum_{k \neq m} 
\left( 
e^{2}_{ik}e^{2}_{jm}
+
e^{2}_{jk}e^{2}_{im}
 \right) 
\gamma_{km}^{\phantom{2}}   
\nonumber \\
&
+4  \sum_{k \neq m}  
e_{ik} \,
e_{jm} \,
e_{jk} \,
e_{im}  \, 
\gamma_{mk}   \quad \mbox{for} \,\,\, i \neq j
 \,  , 
\end{align}
%
%
%
%
with $i,j,k,m = 1,2,3$.
By varying the DC voltage, we modify the hybridization of the eigenmodes
of the systems, namely the DC voltage-dependent components of the eigenvectors $e_{ij}(V_{dc})$.
In consequence,  the Duffing nonlinearity of each mode 
as well as their dispersive coupling to the other modes are altered.
In other words, in the model as given by Eqs.~(\ref{eq_duffing}) and (\ref{eq_dispersive}), 
the Duffing nonlinearities and the dispersive coupling coefficients of the hybridized eigenmodes 
are  voltage-dependent, $\tilde{\gamma}_{ij} (V_{dc})$, 
since they depend on the mode polarization.

Using the theoretical mode polarization discussed in Fig.~\ref{fig3}, we compute the DC-voltage dependence of 
the Duffing nonlinearities $\tilde{\gamma}_{ij}$ 
and of the dispersive coupling coefficients $\tilde{\gamma}_{ij}$.

\begin{figure}[b!]
\centering
\includegraphics[scale=.245]{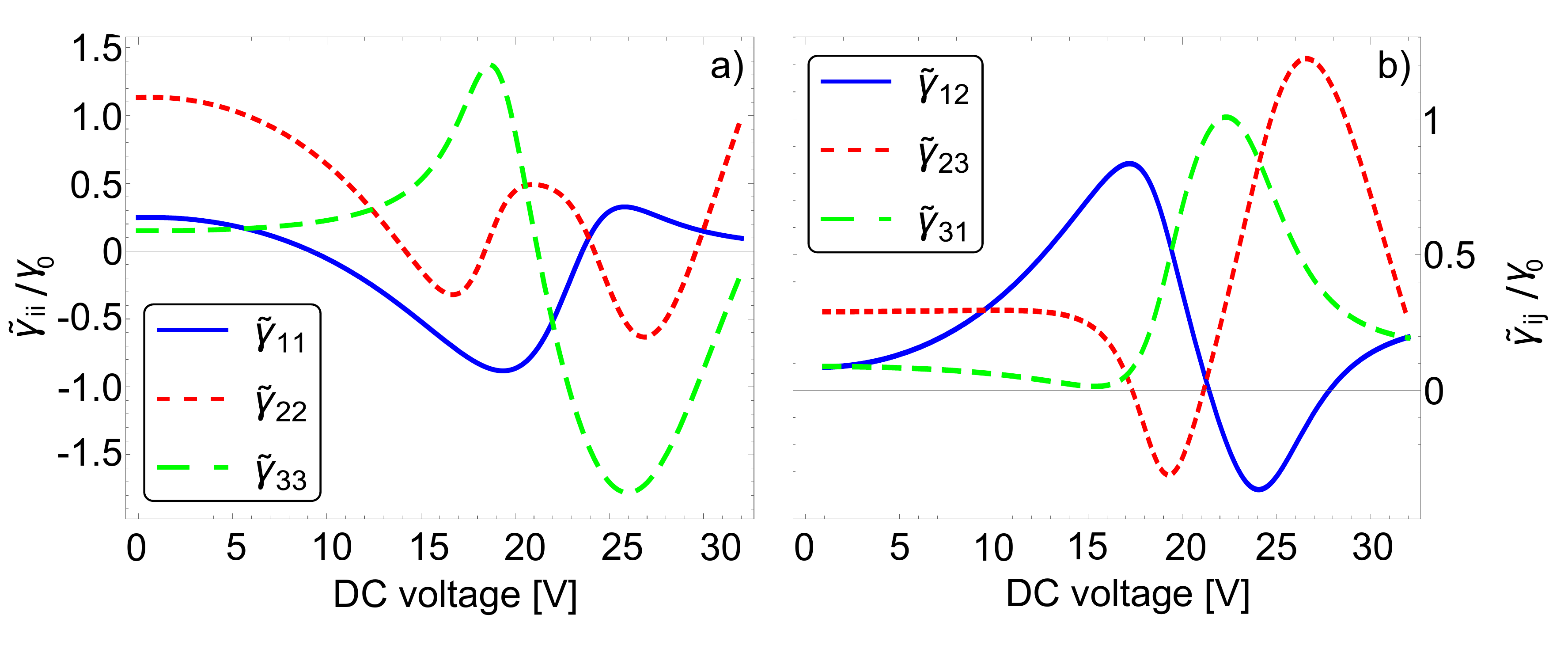}
\caption[]{(color online) 
Theoretical model for the polarization-dependent nonlinear coefficients. a) Duffing nonlinearities $\tilde{\gamma}_{ii}$ ($\tilde{\gamma}_{11}$: blue solid line, $\tilde{\gamma}_{22}$: red short dashed line, $\tilde{\gamma}_{33}$: green long dashed line). 
b) Dispersive coupling coefficients $\tilde{\gamma}_{ij}$ ($\tilde{\gamma}_{12}$: blue solid line, $\tilde{\gamma}_{23}$: red short dashed line, $\tilde{\gamma}_{31}$: green long dashed line).
All plotted quantities have been scaled with respect to $\gamma_0=(\gamma_{11}+\gamma_{22})/6$, and we have used the following realistic 
parameters for the bare modes
$\gamma_{11} /\gamma_0 = 2.2$, $\gamma_{22}/\gamma_0=3.8$,  $\gamma_{33}/\gamma_0=0.01$, and 
$\gamma_{12}/\gamma_0=-6.5$, $\gamma_{13}/\gamma_0=0.8$ $\gamma_{23}/\gamma_0=1.3$.
}
\label{fig6} 
\end{figure}

Figure~\ref{fig6}a and b display the Duffing nonlinearities ($\tilde{\gamma}_{11}$: blue solid line, $\tilde{\gamma}_{22}$: red small dashed line, $\tilde{\gamma}_{33}$: green large dashed line) as well as the dispersive coupling coefficients ($\tilde{\gamma}_{12}$: blue solid line, $\tilde{\gamma}_{23}$: red small dashed line, $\tilde{\gamma}_{31}$: green large dashed line), respectively. It is evident that all nonlinear coefficients strongly depend on the mode hybridization.
%
%
%
%
%
%
%
%
%

At this stage two observations can be made.
First, by varying the DC voltage, we are also varying a priori the intrinsic nonlinear coefficients of the bare modes, namely 
$\gamma_{ij}$ ($i,j = 1,2,3$.).
This effect is not included in the theoretical model given by Eq.~(\ref{eq_duffing}) and Eq.~(\ref{eq_dispersive}) and 
represents an intrinsic property of  dielectrically controlled nanomechanical resonators.
As a second point, we observe that we have measured  the quantities $\tilde{\text{c}}_{i}\tilde{\gamma}_{ij}$
which involve calibration factors $\tilde{\text{c}}_{i}$.
In the present experiment, we have not been able to extract these quantities:
the DC-voltage dependence of the calibration factors $\tilde{\text{c}}_{i}$ is not known. 
Therefore, 
we have opted for a qualitative comparison between theory and experiment  
in order to illustrate the similarities and the differences between the experimental results and the  
expected DC voltage dependence as predicted by the mode polarization model.

However, the order of magnitude for the variations of the experimentally obtained values 
are clearly comparable with the predictions of the theory. 
Moreveover, a strong dependence on the voltage is experimentally observed when the modes are strongly hybridized in the vicinity of the avoided crossings.

\section{Conclusion}
\label{sec:6}
We observed nonlinear dispersive mode coupling in a nanomechanical system formed by three linearly coupled bare eigenmodes of a pair of nanostring resonators.

We also observed a strong voltage dependence of the nonlinear coefficients of the system. Both the Duffing nonlinearities~\cite{Singh:2019} and the dispersive coupling coefficients are modified, particularly in the DC-voltage regime in which strong hybridization occurs.
%
To understand the latter effect, 
we have analyzed  a minimal theoretical model which yields a voltage dependence resulting from the hybridization of the modes.

We showed, that even if a direct comparison between experimental data and theory  was not possible, 
the theoretical model based on the mode hybridization  qualitatively reproduces the non-trivial voltage dependence.
From the experimental side, for further investigations on the voltage dependence of the nonlinear coefficients,
the displacement detection scheme needs to be improved to allow for a full calibration of the measured data.
In addition, a second microwave cavity should be included for resonator 2 to enable the direct detection of all modes.
From the theoretical side, the effects of the voltage dependence of the intrinsic parameters of the systems for the bare modes have to be taken into account 
with a microscopic model for the dielectric modulation of the nonlinearities.
Future work will address these directions.


\begin{acknowledgments}
Financial support from the Zukunftskolleg of the University of Konstanz, the Deutsche Forschungsgemeinschaft
via the Collaborative Research Center SFB 767; the European Unions Horizon 2020 programme for Research
and Innovation under grant agreement No. 732894 (FET Proactive HOT), as well as the German Federal Ministry of Education and Research through contract no. 13N14777 funded within the European QuantERA cofund project QuaSeRT is gratefully acknowledged.
 \end{acknowledgments}


%
%
%
%

\bibliography{dispersive_coupling}

\begin{thebibliography}{16}%
\makeatletter
\providecommand \@ifxundefined [1]{%
 \@ifx{#1\undefined}
}%
\providecommand \@ifnum [1]{%
 \ifnum #1\expandafter \@firstoftwo
 \else \expandafter \@secondoftwo
 \fi
}%
\providecommand \@ifx [1]{%
 \ifx #1\expandafter \@firstoftwo
 \else \expandafter \@secondoftwo
 \fi
}%
\providecommand \natexlab [1]{#1}%
\providecommand \enquote  [1]{``#1''}%
\providecommand \bibnamefont  [1]{#1}%
\providecommand \bibfnamefont [1]{#1}%
\providecommand \citenamefont [1]{#1}%
\providecommand \href@noop [0]{\@secondoftwo}%
\providecommand \href [0]{\begingroup \@sanitize@url \@href}%
\providecommand \@href[1]{\@@startlink{#1}\@@href}%
\providecommand \@@href[1]{\endgroup#1\@@endlink}%
\providecommand \@sanitize@url [0]{\catcode `\\12\catcode `\$12\catcode
  `\&12\catcode `\#12\catcode `\^12\catcode `\_12\catcode `\%12\relax}%
\providecommand \@@startlink[1]{}%
\providecommand \@@endlink[0]{}%
\providecommand \url  [0]{\begingroup\@sanitize@url \@url }%
\providecommand \@url [1]{\endgroup\@href {#1}{\urlprefix }}%
\providecommand \urlprefix  [0]{URL }%
\providecommand \Eprint [0]{\href }%
\providecommand \doibase [0]{http://dx.doi.org/}%
\providecommand \selectlanguage [0]{\@gobble}%
\providecommand \bibinfo  [0]{\@secondoftwo}%
\providecommand \bibfield  [0]{\@secondoftwo}%
\providecommand \translation [1]{[#1]}%
\providecommand \BibitemOpen [0]{}%
\providecommand \bibitemStop [0]{}%
\providecommand \bibitemNoStop [0]{.\EOS\space}%
\providecommand \EOS [0]{\spacefactor3000\relax}%
\providecommand \BibitemShut  [1]{\csname bibitem#1\endcsname}%
\let\auto@bib@innerbib\@empty
\bibitem [{\citenamefont {Mangussi}\ and\ \citenamefont
  {Zanette}(2016)}]{Mangussi2016}%
  \BibitemOpen
  \bibfield  {author} {\bibinfo {author} {\bibfnamefont {F.}~\bibnamefont
  {Mangussi}}\ and\ \bibinfo {author} {\bibfnamefont {D.~H.}\ \bibnamefont
  {Zanette}},\ }\href {\doibase 10.1371/journal.pone.0162365} {\bibfield
  {journal} {\bibinfo  {journal} {{PLOS} {ONE}}\ }\textbf {\bibinfo {volume}
  {11}},\ \bibinfo {pages} {e0162365} (\bibinfo {year} {2016})}\BibitemShut
  {NoStop}%
\bibitem [{\citenamefont {Vinante}(2014)}]{Vinante2014}%
  \BibitemOpen
  \bibfield  {author} {\bibinfo {author} {\bibfnamefont {A.}~\bibnamefont
  {Vinante}},\ }\href {\doibase 10.1103/PhysRevB.90.024308} {\bibfield
  {journal} {\bibinfo  {journal} {Phys. Rev. B}\ }\textbf {\bibinfo {volume}
  {90}},\ \bibinfo {pages} {024308} (\bibinfo {year} {2014})}\BibitemShut
  {NoStop}%
\bibitem [{\citenamefont {Westra}\ \emph {et~al.}(2010)\citenamefont {Westra},
  \citenamefont {Poot}, \citenamefont {van~der Zant},\ and\ \citenamefont
  {Venstra}}]{Westra}%
  \BibitemOpen
  \bibfield  {author} {\bibinfo {author} {\bibfnamefont {H.~J.~R.}\
  \bibnamefont {Westra}}, \bibinfo {author} {\bibfnamefont {M.}~\bibnamefont
  {Poot}}, \bibinfo {author} {\bibfnamefont {H.~S.~J.}\ \bibnamefont {van~der
  Zant}}, \ and\ \bibinfo {author} {\bibfnamefont {W.~J.}\ \bibnamefont
  {Venstra}},\ }\href {\doibase 10.1103/PhysRevLett.105.117205} {\bibfield
  {journal} {\bibinfo  {journal} {Phys. Rev. Lett.}\ }\textbf {\bibinfo
  {volume} {105}},\ \bibinfo {pages} {117205} (\bibinfo {year}
  {2010})}\BibitemShut {NoStop}%
\bibitem [{\citenamefont {van~der Avoort}\ \emph {et~al.}(2010)\citenamefont
  {van~der Avoort}, \citenamefont {van~der Hout}, \citenamefont {Bontemps},
  \citenamefont {Steeneken}, \citenamefont {Phan}, \citenamefont {Fey},
  \citenamefont {Hulshof},\ and\ \citenamefont {van Beek}}]{Avoort2010}%
  \BibitemOpen
  \bibfield  {author} {\bibinfo {author} {\bibfnamefont {C.}~\bibnamefont
  {van~der Avoort}}, \bibinfo {author} {\bibfnamefont {R.}~\bibnamefont
  {van~der Hout}}, \bibinfo {author} {\bibfnamefont {J.~J.~M.}\ \bibnamefont
  {Bontemps}}, \bibinfo {author} {\bibfnamefont {P.~G.}\ \bibnamefont
  {Steeneken}}, \bibinfo {author} {\bibfnamefont {K.~L.}\ \bibnamefont {Phan}},
  \bibinfo {author} {\bibfnamefont {R.~H.~B.}\ \bibnamefont {Fey}}, \bibinfo
  {author} {\bibfnamefont {J.}~\bibnamefont {Hulshof}}, \ and\ \bibinfo
  {author} {\bibfnamefont {J.~T.~M.}\ \bibnamefont {van Beek}},\ }\href
  {\doibase 10.1088/0960-1317/20/10/105012} {\bibfield  {journal} {\bibinfo
  {journal} {J. of Micromech. and Microen.}\ }\textbf {\bibinfo {volume}
  {20}},\ \bibinfo {pages} {105012} (\bibinfo {year} {2010})}\BibitemShut
  {NoStop}%
\bibitem [{\citenamefont {Lulla}\ \emph {et~al.}(2012)\citenamefont {Lulla},
  \citenamefont {Cousins}, \citenamefont {Venkatesan}, \citenamefont {Armour},
  \citenamefont {Mellor},\ and\ \citenamefont {Owers-Bradley}}]{Lulla}%
  \BibitemOpen
  \bibfield  {author} {\bibinfo {author} {\bibfnamefont {K.~J.}\ \bibnamefont
  {Lulla}}, \bibinfo {author} {\bibfnamefont {R.~B.}\ \bibnamefont {Cousins}},
  \bibinfo {author} {\bibfnamefont {M.~J.}\ \bibnamefont {Venkatesan},
  \bibfnamefont {A.~andPatton}}, \bibinfo {author} {\bibfnamefont {A.~D.}\
  \bibnamefont {Armour}}, \bibinfo {author} {\bibfnamefont {C.~J.}\
  \bibnamefont {Mellor}}, \ and\ \bibinfo {author} {\bibfnamefont {J.~R.}\
  \bibnamefont {Owers-Bradley}},\ }\href {\doibase
  10.1088/1367-2630/14/11/113040} {\bibfield  {journal} {\bibinfo  {journal}
  {New J. Phys.}\ }\textbf {\bibinfo {volume} {14}},\ \bibinfo {pages} {113040}
  (\bibinfo {year} {2012})}\BibitemShut {NoStop}%
\bibitem [{\citenamefont {Matheny}\ \emph {et~al.}(2013)\citenamefont
  {Matheny}, \citenamefont {Villanueva}, \citenamefont {Karabalin},
  \citenamefont {Sader},\ and\ \citenamefont {Roukes}}]{Matheny}%
  \BibitemOpen
  \bibfield  {author} {\bibinfo {author} {\bibfnamefont {M.~H.}\ \bibnamefont
  {Matheny}}, \bibinfo {author} {\bibfnamefont {L.~G.}\ \bibnamefont
  {Villanueva}}, \bibinfo {author} {\bibfnamefont {R.~B.}\ \bibnamefont
  {Karabalin}}, \bibinfo {author} {\bibfnamefont {J.~E.}\ \bibnamefont
  {Sader}}, \ and\ \bibinfo {author} {\bibfnamefont {M.~L.}\ \bibnamefont
  {Roukes}},\ }\href {\doibase 10.1021/nl400070e} {\bibfield  {journal}
  {\bibinfo  {journal} {Nano Lett.}\ }\textbf {\bibinfo {volume} {13}},\
  \bibinfo {pages} {1622} (\bibinfo {year} {2013})}\BibitemShut {NoStop}%
\bibitem [{\citenamefont {Cadeddu}\ \emph {et~al.}(2016)\citenamefont
  {Cadeddu}, \citenamefont {Braakman}, \citenamefont
  {T{\"{u}}t{\"{u}}nc{\"{u}}oglu}, \citenamefont {Matteini}, \citenamefont
  {R{\"{u}}ffer}, \citenamefont {Fontcuberta~i Morral},\ and\ \citenamefont
  {Poggio}}]{Cadeddu2016}%
  \BibitemOpen
  \bibfield  {author} {\bibinfo {author} {\bibfnamefont {D.}~\bibnamefont
  {Cadeddu}}, \bibinfo {author} {\bibfnamefont {F.~R.}\ \bibnamefont
  {Braakman}}, \bibinfo {author} {\bibfnamefont {G.}~\bibnamefont
  {T{\"{u}}t{\"{u}}nc{\"{u}}oglu}}, \bibinfo {author} {\bibfnamefont
  {F.}~\bibnamefont {Matteini}}, \bibinfo {author} {\bibfnamefont
  {D.}~\bibnamefont {R{\"{u}}ffer}}, \bibinfo {author} {\bibfnamefont
  {A.}~\bibnamefont {Fontcuberta~i Morral}}, \ and\ \bibinfo {author}
  {\bibfnamefont {M.}~\bibnamefont {Poggio}},\ }\href {\doibase
  10.1021/acs.nanolett.5b03822} {\bibfield  {journal} {\bibinfo  {journal}
  {Nano Lett.}\ }\textbf {\bibinfo {volume} {16}},\ \bibinfo {pages} {926}
  (\bibinfo {year} {2016})}\BibitemShut {NoStop}%
\bibitem [{\citenamefont {Dong}\ \emph {et~al.}(2018)\citenamefont {Dong},
  \citenamefont {Dykman},\ and\ \citenamefont {Chan}}]{Dong2018}%
  \BibitemOpen
  \bibfield  {author} {\bibinfo {author} {\bibfnamefont {X.}~\bibnamefont
  {Dong}}, \bibinfo {author} {\bibfnamefont {M.~I.}\ \bibnamefont {Dykman}}, \
  and\ \bibinfo {author} {\bibfnamefont {H.~B.}\ \bibnamefont {Chan}},\ }\href
  {\doibase 10.1038/s41467-018-05246-w} {\bibfield  {journal} {\bibinfo
  {journal} {Nat. Comm.}\ }\textbf {\bibinfo {volume} {9}},\ \bibinfo {pages}
  {3241} (\bibinfo {year} {2018})}\BibitemShut {NoStop}%
\bibitem [{\citenamefont {Mathew}\ \emph {et~al.}(2018)\citenamefont {Mathew},
  \citenamefont {Bhushan},\ and\ \citenamefont {Deshmukh}}]{Mathew2018}%
  \BibitemOpen
  \bibfield  {author} {\bibinfo {author} {\bibfnamefont {J.~P.}\ \bibnamefont
  {Mathew}}, \bibinfo {author} {\bibfnamefont {A.}~\bibnamefont {Bhushan}}, \
  and\ \bibinfo {author} {\bibfnamefont {M.~M.}\ \bibnamefont {Deshmukh}},\
  }\href {\doibase 10.1016/j.ssc.2018.07.009} {\bibfield  {journal} {\bibinfo
  {journal} {Sol. State Comm.}\ }\textbf {\bibinfo {volume} {282}},\ \bibinfo
  {pages} {17} (\bibinfo {year} {2018})}\BibitemShut {NoStop}%
\bibitem [{\citenamefont {Gajo}\ \emph {et~al.}(2017)\citenamefont {Gajo},
  \citenamefont {Sch{\"u}z},\ and\ \citenamefont {Weig}}]{Gajo}%
  \BibitemOpen
  \bibfield  {author} {\bibinfo {author} {\bibfnamefont {K.}~\bibnamefont
  {Gajo}}, \bibinfo {author} {\bibfnamefont {S.}~\bibnamefont {Sch{\"u}z}}, \
  and\ \bibinfo {author} {\bibfnamefont {E.~M.}\ \bibnamefont {Weig}},\ }\href
  {\doibase 10.1063/1.4995230} {\bibfield  {journal} {\bibinfo  {journal}
  {Appl. Phys. Lett.}\ }\textbf {\bibinfo {volume} {111}},\ \bibinfo {pages}
  {133109} (\bibinfo {year} {2017})}\BibitemShut {NoStop}%
\bibitem [{\citenamefont {Unterreithmeier}\ \emph {et~al.}(2009)\citenamefont
  {Unterreithmeier}, \citenamefont {Weig},\ and\ \citenamefont
  {Kotthaus}}]{Unterreithmeier_actuation}%
  \BibitemOpen
  \bibfield  {author} {\bibinfo {author} {\bibfnamefont {Q.~P.}\ \bibnamefont
  {Unterreithmeier}}, \bibinfo {author} {\bibfnamefont {E.~M.}\ \bibnamefont
  {Weig}}, \ and\ \bibinfo {author} {\bibfnamefont {J.~P.}\ \bibnamefont
  {Kotthaus}},\ }\href {\doibase 10.1038/ncomms1723} {\bibfield  {journal}
  {\bibinfo  {journal} {Nature}\ }\textbf {\bibinfo {volume} {458}},\ \bibinfo
  {pages} {1001} (\bibinfo {year} {2009})}\BibitemShut {NoStop}%
\bibitem [{\citenamefont {{Rieger}}\ \emph {et~al.}(2012)\citenamefont
  {{Rieger}}, \citenamefont {{Faust}}, \citenamefont {{Seitner}}, \citenamefont
  {{Kotthaus}},\ and\ \citenamefont {{Weig}}}]{Rieger_tuning}%
  \BibitemOpen
  \bibfield  {author} {\bibinfo {author} {\bibfnamefont {J.}~\bibnamefont
  {{Rieger}}}, \bibinfo {author} {\bibfnamefont {T.}~\bibnamefont {{Faust}}},
  \bibinfo {author} {\bibfnamefont {M.~J.}\ \bibnamefont {{Seitner}}}, \bibinfo
  {author} {\bibfnamefont {J.~P.}\ \bibnamefont {{Kotthaus}}}, \ and\ \bibinfo
  {author} {\bibfnamefont {E.~M.}\ \bibnamefont {{Weig}}},\ }\href {\doibase
  10.1063/1.4751351} {\bibfield  {journal} {\bibinfo  {journal} {Appl. Phy.
  Lett.}\ }\textbf {\bibinfo {volume} {101}},\ \bibinfo {pages} {103110}
  (\bibinfo {year} {2012})}\BibitemShut {NoStop}%
\bibitem [{\citenamefont {Faust}\ \emph {et~al.}(2012)\citenamefont {Faust},
  \citenamefont {Krenn}, \citenamefont {Manus}, \citenamefont {Kotthaus},\ and\
  \citenamefont {Weig}}]{Faust_Mikrowelle}%
  \BibitemOpen
  \bibfield  {author} {\bibinfo {author} {\bibfnamefont {T.}~\bibnamefont
  {Faust}}, \bibinfo {author} {\bibfnamefont {P.}~\bibnamefont {Krenn}},
  \bibinfo {author} {\bibfnamefont {S.}~\bibnamefont {Manus}}, \bibinfo
  {author} {\bibfnamefont {J.~P.}\ \bibnamefont {Kotthaus}}, \ and\ \bibinfo
  {author} {\bibfnamefont {E.~M.}\ \bibnamefont {Weig}},\ }\href {\doibase
  10.1038/ncomms1723} {\bibfield  {journal} {\bibinfo  {journal} {Nat. Comm.}\
  }\textbf {\bibinfo {volume} {3}},\ \bibinfo {pages} {728} (\bibinfo {year}
  {2012})}\BibitemShut {NoStop}%
\bibitem [{\citenamefont {{Huber}}\ \emph {et~al.}(2019)\citenamefont
  {{Huber}}, \citenamefont {{Rastelli}}, \citenamefont {{Seitner}},
  \citenamefont {{K{\"o}lbl}}, \citenamefont {{Belzig}}, \citenamefont
  {{Dykman}},\ and\ \citenamefont {{Weig}}}]{Huber2019}%
  \BibitemOpen
  \bibfield  {author} {\bibinfo {author} {\bibfnamefont {J.~S.}\ \bibnamefont
  {{Huber}}}, \bibinfo {author} {\bibfnamefont {G.}~\bibnamefont {{Rastelli}}},
  \bibinfo {author} {\bibfnamefont {M.~J.}\ \bibnamefont {{Seitner}}}, \bibinfo
  {author} {\bibfnamefont {J.}~\bibnamefont {{K{\"o}lbl}}}, \bibinfo {author}
  {\bibfnamefont {W.}~\bibnamefont {{Belzig}}}, \bibinfo {author}
  {\bibfnamefont {M.~I.}\ \bibnamefont {{Dykman}}}, \ and\ \bibinfo {author}
  {\bibfnamefont {E.~M.}\ \bibnamefont {{Weig}}},\ }\href@noop {} {\bibfield
  {journal} {\bibinfo  {journal} {e-prints}\ ,\ \bibinfo {eid}
  {arXiv:1903.07601}} (\bibinfo {year} {2019})},\ \Eprint
  {http://arxiv.org/abs/1903.07601} {1903.07601} \BibitemShut {NoStop}%
\bibitem [{\citenamefont {Zhang}\ and\ \citenamefont
  {Dykman}(2015)}]{Zhang2015}%
  \BibitemOpen
  \bibfield  {author} {\bibinfo {author} {\bibfnamefont {Y.}~\bibnamefont
  {Zhang}}\ and\ \bibinfo {author} {\bibfnamefont {M.~I.}\ \bibnamefont
  {Dykman}},\ }\href {\doibase 10.1103/PhysRevB.92.165419} {\bibfield
  {journal} {\bibinfo  {journal} {Phys. Rev. B}\ }\textbf {\bibinfo {volume}
  {92}},\ \bibinfo {pages} {165419} (\bibinfo {year} {2015})}\BibitemShut
  {NoStop}%
\bibitem [{\citenamefont {Singh}\ \emph {et~al.}(2019)\citenamefont {Singh},
  \citenamefont {Sarkar}, \citenamefont {Guria}, \citenamefont {Nicholl},
  \citenamefont {Chakraborty}, \citenamefont {Bolotin},\ and\ \citenamefont
  {Ghosh}}]{Singh:2019}%
  \BibitemOpen
  \bibfield  {author} {\bibinfo {author} {\bibfnamefont {R.}~\bibnamefont
  {Singh}}, \bibinfo {author} {\bibfnamefont {A.}~\bibnamefont {Sarkar}},
  \bibinfo {author} {\bibfnamefont {C.}~\bibnamefont {Guria}}, \bibinfo
  {author} {\bibfnamefont {R.}~\bibnamefont {Nicholl}}, \bibinfo {author}
  {\bibfnamefont {S.}~\bibnamefont {Chakraborty}}, \bibinfo {author}
  {\bibfnamefont {K.~I.}\ \bibnamefont {Bolotin}}, \ and\ \bibinfo {author}
  {\bibfnamefont {S.}~\bibnamefont {Ghosh}},\ }\href@noop {} {\bibfield
  {journal} {\bibinfo  {journal} {eprint arXiv:1904.01613}\ } (\bibinfo {year}
  {2019})}\BibitemShut {NoStop}%
\end{thebibliography}%

\end{document}